\documentclass[pre,showpac,preprintnumbers,twocolumn]{revtex4}
\usepackage{epsfig}
\usepackage{amssymb,amsmath,amsthm}
\def \ls{localized structures }
\def \ss{solitary structures }

\begin{document}
\title{Pattern collapse as a mechanism for the formation of solitary structures}

\author{U. Bortolozzo$^{1}$, M.G. Clerc$^{2}$, C. Falc\'on$^{1}$, S. Residori$^{3}$}

\address{$^{1}$Laboratoire de Physique Statistique de l'Ecole Normale Sup\'erieure, 24 rue Lhomond, 75231 Paris cedex 05, France
\\
$^{2}$Departamento de F\'{i}sica, Universidad de Chile, Casilla 487-3, Santiago, Chile.
\\
$^{3}$INLN, Universit\'e de Nice Sophia-Antipolis, CNRS, 1361 route des Lucioles, 06560 Valbonne, France
 }

\date{\today}

\begin{abstract}
We report a new mechanism for the  formation of localized states, which takes place without front propagation. Correspondingly, localized structures appear as solitary states, displaying a behavior of single independent cells. The phenomenon is observed in the liquid crystal light-valve experiment and is described by a one-dimensional normal form model. We show that such solitary structures exist when a pattern solution collapses and its ghost remains to influence the phase portrait. 
\end{abstract}


\maketitle

The concept of {\it soliton} as a single coherent structure
originates  from Hamiltonian systems \cite{Newell}. The
generalization of this concept to dissipative and
out of equilibrium systems has led to 
several studies in the last decades and to the definition of a class of {\it localized structures} that are intended as patterns appearing in a restricted region of space. Localized structures have been observed in different fields, such as in magnetic materials, chemical reactions, granular media, plasmas, nonlinear optics and
surface waves \cite{Cross}. Theoretically, they are understood as macroscopic particle-like objects arising in the presence of metastability between different states. Localized structures, thus, realize the
spatial connection between two possible states and appear in the \textit{pinning range} of the front solution \cite{Pomeau}. 
In one-dimensional systems and from the geometrical point of view, they are described by homoclinic orbits passing close to a pattern
state and converging to an homogeneous state
\cite{VanSaarlos,Coullet2000}. More recently, they have been explained in terms of front interactions \cite{ClercFalcon2005} and their existence has been generalized to the case when the homoclinic orbits connect two different pattern states, thus leading to \textit{localized peaks} \cite{notrePRL}.

In all these theoretical frameworks, \ls are predicted to exist in parameter regions where the system exhibits coexistence between two spatially extended states and are expected to
appear (disappear) by a sequence of saddle-node
bifurcations, each bifurcation leading to a larger (smaller) number of pattern cells and occurring in the pinning range of the {\it Pomeau front} \cite{Pomeau}.
Nevertheless, there is a wealth of experimental observations, for example in optics \cite{Lange,Ramazza} or in vibrated granular media \cite{Melo}, of solitary states that display more a character of single objects than that of a class of differently sized patterns, that is, only the single-cell localized structures appear. Recent numerical
simulations have also shown that cavity solitons occurring in semiconductor microcavities have their region of existence outside the range of bistability between the homogeneous and the pattern states \cite{Tissoni},
which is consistent with the experimental observations
\cite{InlnSemiconductorLaser}. The common property of all these
observations is that \ls  seem to exist without the need of metastability, which claims for searching
new mechanisms for their formation.

The aim of this letter is to show a new type of transition from extended patterns to localized structures, taking place without front propagation and leading to {\it solitary structures} that appear randomly in space, behave as independent particles and interact through the oscillations on their tails. We explain this transition through the collapse of the pattern solution via a saddle-node bifurcation, which implies the destruction of the homoclinic snaking sequence associated to the whole family of differently sized localized patterns. After the collapse, the ghost of the pattern remains to influence the phase portrait and solitary (one-cell) structures exist as the result of the unique (single-loop) surviving homoclinic orbit. 
We present experimental evidence of this transition for a liquid crystal light-valve (LCLV) with optical feedback. The results are confirmed by numerical simulations of the full model. Then, we introduce a 1D normal form equation, which allows the use of general geometrical arguments to describe the pattern collapse.

\begin{figure}[h!]
\epsfclipon \epsfig{file=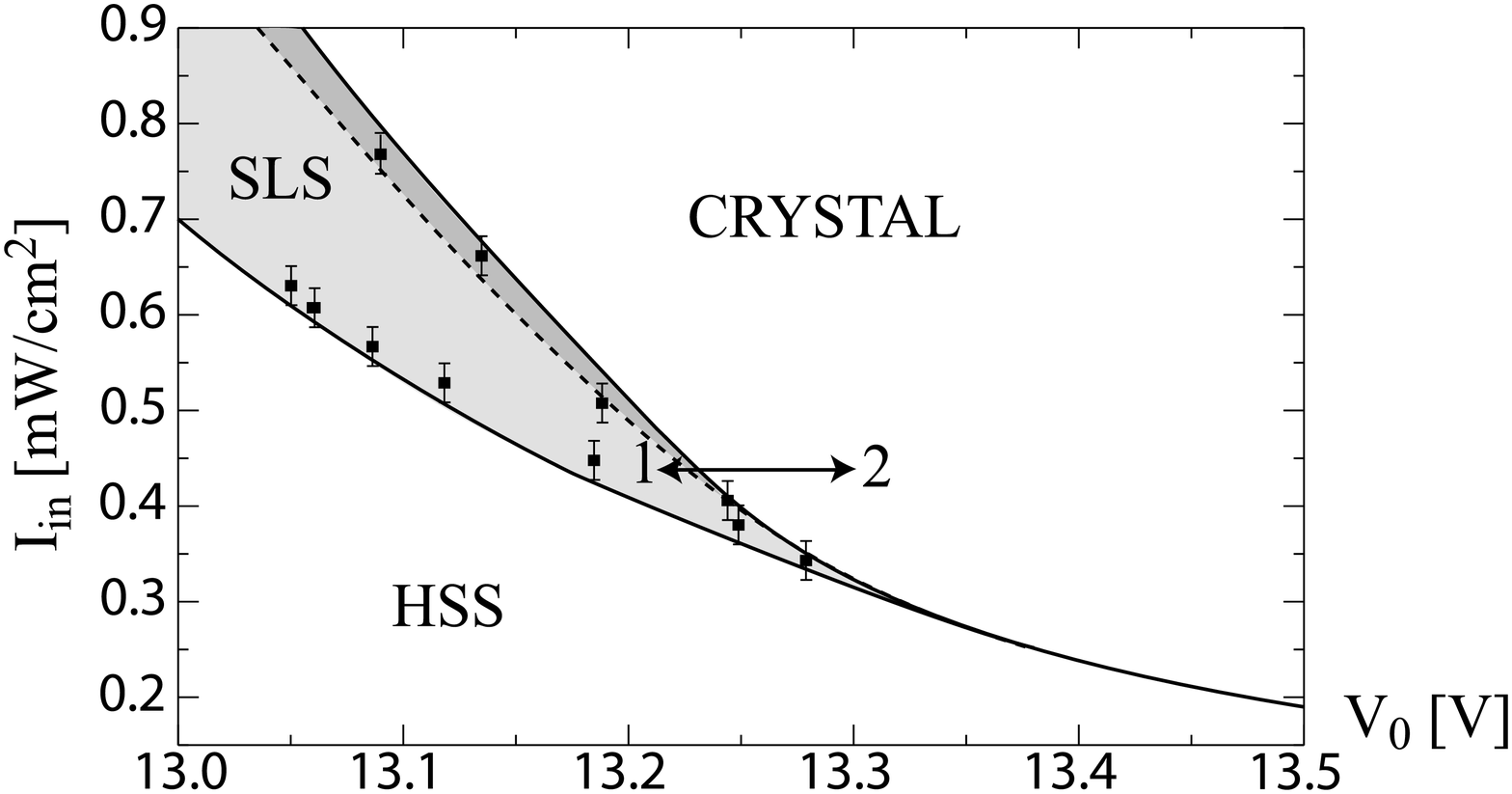,width=7.5 cm}
\caption{Phase diagram $I_{in}$ vs $V_0$.
Continuos lines correspond to the
numerically calculated boundaries from crystal to solitary structures and to homogeneous state. The dashed line delimits the region of bistability between the homogeneous and the crystal state. Dots are experimental points.
\label{experiment1}}
\end{figure}

The experiment consists of a LCLV inserted in an optical feedback loop \cite{Residori-report}. When a voltage $V_0$ is applied to the valve, the voltage that effectively drops across the liquid crystals is $V_{LC}= \Gamma V_0 + \alpha I_w$, where $I_w$ is the light intensity impinging on the photoconductor side of the LCLV  and $\Gamma$, $\alpha$ are phenomenological parameters summarizing, in the linear approximation, the response of the photoconductor. 
The input beam, when passing through the liquid crystal layer, undergoes a phase shift $\varphi= \beta  \cos^2 \theta$, with $\theta$ the average tilt angle of the liquid crystal molecules and
$\beta=2 \pi \Delta n d / \lambda$, where $\lambda$ is the optical wavelength
and $\Delta n=n_e-n_0$ is the liquid crystal birefringence, $n_e$ and $n_o$ being, respectively, the extraordinary (parallel to the liquid crystal director 
$\vec n$) and ordinary (perpendicular to $\vec n$) refractive index. 
The tilt angle $\theta$ obeys a Debye relaxation equation \cite{duemarroni}, 

\begin{equation}
\tau \partial _{t}\theta =l^{2}\nabla_\perp^2\theta -\theta + f(\theta)
\label{eq1}
\end{equation}
where $l=20$ $\mu m$ is the diffusion length summarizing the elastic coupling in the liquid crystal and the charge diffusion in the photoconductor, $\tau=10$ $ms$ is the local relaxation time, $V_{FT}=1.05$ $V$ is the Fr\'eedericksz transition voltage and $f(\theta)=\pi /2 \left(1- \sqrt{V_{FT} / V_{LC}} \right)$ accounts for the response of the LCLV \cite{notrePRL1}.
The optical feedback is obtained by sending back onto the photoconductor the light that has passed through the liquid crystals and has been reflected by the LCLV. After free propagation and polarization interference, the light intensity arriving at the photoconductor is 
\begin{equation}
I_{w}= {I_{in} \over 4}  \mid e^{i {L \lambda \over 4 \pi} \nabla_\perp^2} 
\cdot \left( 1- e^{ i \beta \cos^2 \theta} \right)  \mid^2,
\label{eq2}
\end{equation}
where $I_{in}$ is the input intensity and $L$ is the free propagation length in the optical feedback loop.

\begin{figure}[h!]
\epsfclipon \epsfig{file=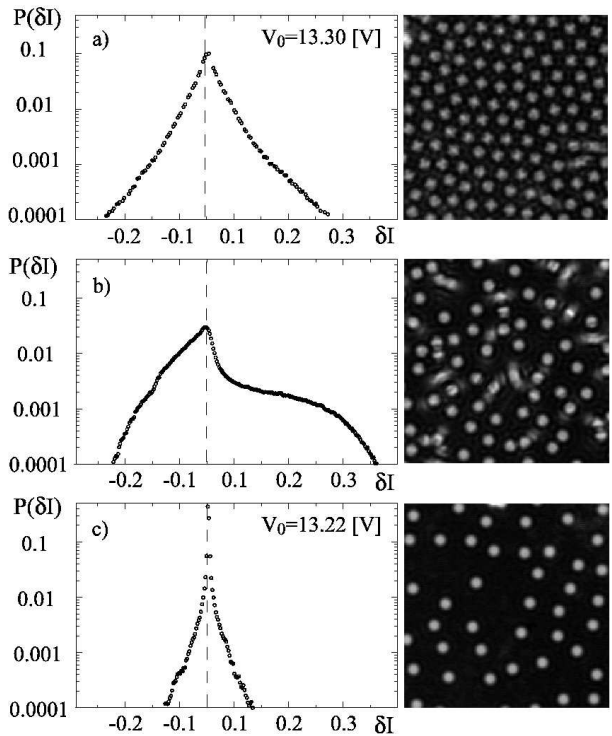,width=8cm}
\caption{Experimental PDF (left) and snapshots (right) showing the transition from a) an extended pattern to c) a frozen configuration of solitary structures, through b)  their interaction and annihilation.
\label{experiment}}
\end{figure}

In the experiment, we have fixed the free propagation length to $L=8$ $cm$ and we change the applied voltage $V_0$ and the input light intensity $I_{in}$ as control parameters. By varying either $V_0$ or $I_{in}$ we observe different regimes of patterns and localized structures. The transition from an extended crystal-like pattern to \ss occur in a wide range of the $V_0$, $I_{in}$ parameters.
Fig.\ref{experiment1} shows the region of existence of \ss in the $V_0-I_{in}$ plane. Dots are experimental points obtained by detecting the pattern change when varying the voltage $V_0$ for each value of the pump intensity $I_{in}$. Lines are numerically calculated from the model equations Eqs.(\ref{eq1}) and (\ref{eq2}), in the following way. We have fixed the parameters in the region of existence of solitary localized structures ($SLS$), then 
a single structure is switched on through a light pulse. The voltage is increased until the transition
toward hexagons occurs or is decreased until the homogeneous steady state ($HSS$) is reached. The same procedure is repeated starting from the extended pattern and by decreasing the voltage until the homogeneous solution is approched. The continuos lines marks  the boundaries of the $SLS$ region. The dashed line delimits the region of bistability between the homogeneous and the pattern state.
This region is very thin and is not observed in the experiment due to liquid crystal inhomogeneities.
Note also that for input intensity $I_{in} \ge 0.8$ $mW/cm^2$ a transition from hexagons to
space-time chaos occurs and for voltage $V_0 \le 13.05$ $V$ we observe the formation of triangular localized structures \cite{triangolo}. In this case the transition to extended pattern shows a more 
complex scenario, that will not be reported here.

Experimentally, the boundary between $HSS$ and $SLS$ is found by applying a local perturbation at the $HSS$ state and by checking if  one or more solitary structures remain after removing the perturbation, while the boundary between the crystal and the $SLS$ is determined by starting with $V_0$ in the crystal region and then decreasing it to the $SLS$ region.
We show in Fig.\ref{experiment} three typical snapshots recorded 
for $I_{in}=0.38$ $mW/cm^2$ and by changing the $rms$ value of $V_0$ from $13.30$ to $13.22$ $V$. We first observe an hexagonal pattern (Fig.\ref{experiment}a). When decreasing the voltage, a transition takes place to a final distribution of solitary structures appearing in random space positions (Fig.\ref{experiment}c). 
During the transition solitary single-cell structures interact one with each other through the low amplitude oscillations around their profile (Fig.\ref{experiment}b).
This state displays a gas-like behavior, in the sense that it is mainly characterized by continuous collisions between the particles. However, at difference with a real gas, the collisions may lead also to particle annihilation, so that during the time the total number of particles is not conserved and their mean distance increases.  After a transient, which lasts for a few seconds, a final frozen configuration of particles is reached, where \ss remain fixed in their position. 
Starting from different initial conditions different final frozen configurations are observed, so that we can identify the state of Fig.\ref{experiment}c as one with a large configurational entropy, in the sense defined in Ref.\cite{Coullet-Toniolo}. 
Note that \ss always remain individuals and do not form a pattern, as can be remarked by the empty space between them. 
We can also note in Fig.\ref{experiment}b (at difference with Fig.\ref{experiment}a) the presence of oscillatory rings around each individual cell, these rings being
due to the diffraction of a single spot over the uniform background and, thus, representing a strong signature of the absence of front propagation during the transition.

\begin{figure}[h!]
\epsfclipon \epsfig{file=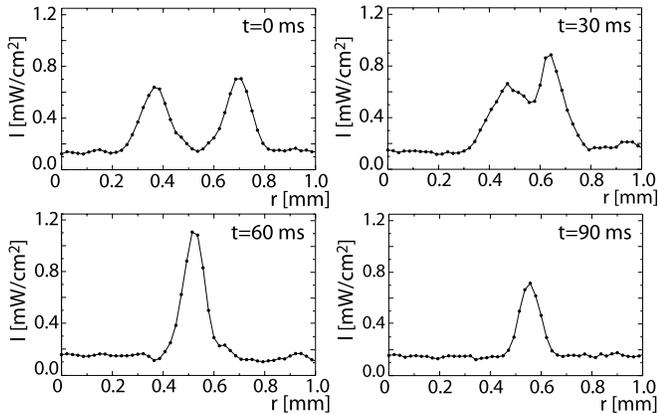,width=\columnwidth}
\caption{Experimental profiles recorded during a two-particle collision: after the collision one particle has been annihilated.
\label{collision}}
\end{figure}

In order to characterize the different states, we have recorded long-time movies and from each movie we have extracted the probability distribution function (PDF) of the light intensity fluctuations $\delta I=I-<I>$, where I is the intensity of each pixel and $<I>$ is the average intensity distribution, which is calculated by averaging pixel by pixel over the entire stack of images. The results are shown in the left part of Fig.\ref{experiment}. The crystal and the frozen configuration of \ss are almost stationary states and the PDFs are given by fluctuations of the particle around their equilibrium positions. In the crystal case the PDF is larger because of slow sliding and gliding of differently oriented domains, whereas for the frozen gas only the fluctuations due to inhomogeneities and noise naturally present in the system are influencing the stationary state.
The transition from hexagons to solitary structures is characterized by an abrupt change of the PDF that becomes strongly asymmetric acquiring a high tail at large intensity fluctuations (Fig.\ref{experiment}b). These large fluctuations correspond to events of particle interactions, when high intensity pulses are produced during the collisions of two particles.
In Fig.\ref{collision} is displayed a set of experimental profiles recorded during an event of two-particle collision. It can be seen that a high intensity pulse is produced when the two particles collide and that after the collision one particle has been annihilated.

\begin{figure}[h!]
\epsfclipon \epsfig{file=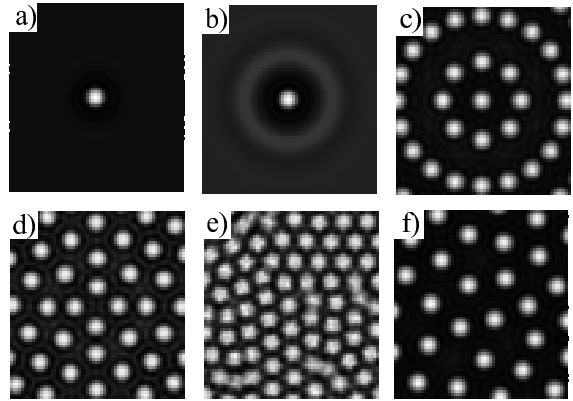,width=7.5cm}
\caption{Numerical snapshots showing the evolution of the  liquid crystal tilt angle $\theta(x,y,t)$ during the transition $1 \rightarrow 2 \rightarrow 1$ indicated in Fig.\ref{experiment1}, for $I_{in}= 0.45$ $mW/cm^2$. a) $V_0=13.30$ $V$ (state $1$); b),c),d),e) time-sequence to reach the state $2$, $V_0=13.22$ $V$; f) state $1$, same values of $V_0$ as in a).
\label{numerics}}
\end{figure}

Fig.\ref{numerics} shows numerical snapshots of liquid crystal tilt angle $\theta(x,y,t)$ during the transition $1 \rightarrow 2 \rightarrow 1$, as marked in Fig.\ref{experiment1}. In Fig.\ref{numerics}a a single cell structure is switched on by a triggering intensity pulse (state $1$ in Fig.\ref{experiment1}). Then, the voltage is increased to reach the crystalline structure of Fig.\ref{numerics}e (state $2$). Figs.\ref{numerics}b,c,d show the 
transient towards the crystal state.
During this transition we observe that new cells nucleate spontaneously around the starting one. In the experiment this effect is not observed because of the presence of spatial inhomogeneities in the LCLV, thus the structures are spontaneously created in random space positions. 
When the voltage is decreased back to its initial value, state $1$, an ensemble of \ss is obtained, as displayed in Fig.\ref{numerics}f.
The transition from crystal to \ss displays the same scenario as the one observed in the experiment, with a transient characterized by particle interaction. Correspondingly, the PDF of the  intensity obtained from the numerical simulations are in good agreement with the experimental PDFs.

To describe a possible mechanism for the formation of solitary structures we consider a simple one-dimensional model based on a Ginzburg-Landau equation with spatial forcing \cite{notrePRL}.
The required ingredients of the model are the coexistence between an uniform and a pattern
state and a saddle-node bifurcation leading to the collapse of the pattern solution under the variation of a control parameter.
We take the following normal form
\begin{equation}
\partial_{t}A=\varepsilon A-\nu|A|^{2}A+\alpha|A|^{4}A
+|A|^{6}A+\partial_{xx}A+\eta A^{2}e^{iqx},
\label{model}
\end{equation}
which is a generic one and can be shown to apply to the case of the LCLV for the range of parameters corresponding to the region of observation of solitary structures.
$A(x,t)$ is a complex amplitude, $\epsilon$ is the bifurcation parameter,$(\nu,\alpha)$ control the type of the bifurcation, $\eta$ is the intensity of the spatial forcing and $q$ is the wave number of the spatially periodic forcing.
For negative $\varepsilon$, $\nu$, $\alpha$ and small $\eta$, the
system has a stable uniform state, $|A|=0$, plus one stable and two unstable pattern states. 
Numerically, we have obtained solitary structures when the stable and unstable pattern 
collapse through a saddle-node bifurcation. 
In correspondence, all patterns collapse to the smallest (single-cell) one and only solitary structures are observed, which continue to exist in a large region outside the region of bistability 
between the homogeneous and the pattern state.

\begin{figure}[h!]
\epsfclipon \epsfig{file=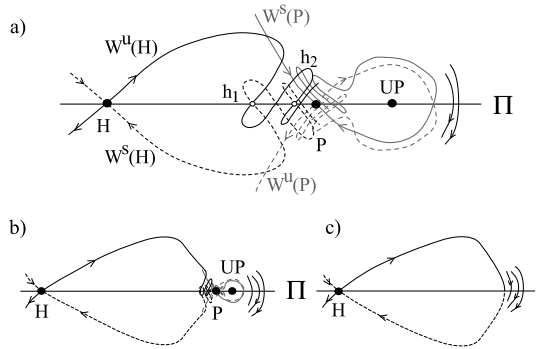,width=7cm}
\caption{Evolution of the homoclinic structure in the phase portrait of the Poincar\'e section: a) the homoclinic snaking  occurring in the pinning range and far from the saddle-node bifurcation ($\eta=0.3$,$\epsilon=-2.6$), b) close ($\eta=0.3$,$\epsilon=-2.5$) and c) after the saddle-node bifurcation ($\eta=0.3$,$\epsilon=-2.4$). (The other parameters are $\nu=0.7$,
$\alpha=-4.9$, $q=12$).}
\label{Fig-HomoclinicSnaking}
\end{figure}

To illustrate the phenomenon of the pattern collapse, we show in  Fig.\ref{Fig-HomoclinicSnaking} how it is modified the typical scenario of homoclinic snaking occurring
in the Poincar\'{e} section of the spatial reversal stationary dynamical system \cite{Champneys}. Fig.\ref{Fig-HomoclinicSnaking}a represents the pinning range. $\Pi$ is
the symmetry plane related to the spatial reflection symmetry ($x\rightarrow -x$), that is, the trajectories are symmetrical with respect to this plane. For the model Eq.(\ref{model}) the reflection symmetry has the form $x\rightarrow-x$ and $A\rightarrow\bar{A}$, the symmetry plane is defined by $\Pi=\{\Re(\partial_{x}A)=\Im(A)=0\}$. The points $H$, $P$, and $UP$ represent, respectively, the stable homogeneous solution, the stable pattern solution and the unstable pattern solution. The continuous and dashed dark $\left\{  W^{s}(H),W^{u}(H)\right\}$ and grey $\left\{ W^{s}(P),W^{u}(P)\right\}$ curves are the stable and unstable manifold of $H$ and $P$, respectively. The homoclinic curves, represented by the points $\left\{h_{1},h_{2},\cdots\right\}$ correspond to the differently sized localized patterns with one, two, three cells and so forth, in agreement with the bifurcation diagram of localized patterns proposed in \cite{Coullet2000}.
When, by changing a control parameter, $UP$ approaches $P$, the influence of the heteroclinic tangle is decreased, as illustrated in Fig.\ref{Fig-HomoclinicSnaking}b. After the saddle node bifurcation the pattern solution does not exists any more and all the homoclinic solutions collapse in the same one, as depicted in Fig.
\ref{Fig-HomoclinicSnaking}c. The saddle-node bifurcation continues to influence the manifold evolution, because the
pattern leaves its ghost in the phase space, which continues to attracts the phase space trajectories \cite{Strogatz}. As a consequence, the remaining homoclinic orbit represents a solitary (single-cell) localized structure, which, by topological reasons, continues to survive until it collides with another homoclinic solution. Note that a similar mechanism has been recently proposed to explain the transition from oscillating \ls to an excitability regime, where the collapse takes place through the merging of a limit cycle with a saddle point \cite{Colet}.

In conclusion, we have shown that pattern collapse is a new mechanism for the formation of solitary structures, the minimal requirement being the presence of a stable solution with a pattern ghost that allows for the existence of a single-loop homoclinic orbit. 

The authors thanks the support of ECOS-CONICYT collaboration program. M.G. C. acknowledges the financial support from the  ring program ACT15 of Programa Bicentenario. U. Bortolozzo acknowledges a fellowship of the {\it Ville de Paris}.

\end{document}